\documentclass[aps,prl,twocolumn,groupedaddress]{revtex4}
\usepackage{graphicx}

\begin{document}


\title{Ion Irradiation Control of Ferromagnetism in (Ga,Mn)As}



\author{H. Kato}
\author{K. Hamaya}
\email{hamaya@iis.u-tokyo.ac.jp}
\thanks{Present adress: Institute of Industrial Science, The University of Tokyo, 4-6-1 Komaba, Meguro-ku, Tokyo 153-8505, Japan.}
\author{T. Taniyama$^{1}$}
\author{Y. Kitamoto}
\author{H. Munekata$^{2}$}

\affiliation{%
Department of Innovative and Engineered Materials, Tokyo Institute of
Technology,\\ 4259 Nagatsuta, Midori-ku, Yokohama 226-8502, Japan.\\
$^{1}$Materials and Structures Laboratory, Tokyo Institute of Technology,\\ 4259 Nagatsuta, Midori-ku, Yokohama 226-8503, Japan.\\
$^{2}$Imaging Science and Engineering Laboratory, Tokyo Institute of
Technology,\\ 4259 Nagatsuta, Midori-ku, Yokohama 226-8503, Japan.
}%



\date{\today}
\begin{abstract}
We report on a promising approach to the artificial modification of ferromagnetic properties in (Ga,Mn)As using a Ga$^+$ focused ion beam (FIB) technique. The ferromagnetic properties of (Ga,Mn)As such as magnetic anisotropy and Curie temperature can be controlled using Ga$^+$ ion irradiation, originating from a change in hole concentration and the corresponding systematic variation in exchange interaction between Mn spins. This change in hole concentration is also verified using micro-Raman spectroscopy. We envisage that this approach offers a means of modifying the ferromagnetic properties of magnetic semiconductors on the micro- or nano-meter scale.

\end{abstract}


\maketitle

Extensive studies of the ferromagnetic properties of $p$-type Mn-doped III-V semiconductors (In,Mn)As and (Ga,Mn)As have been heralding a new era of spin electronics in recent years \cite{Koshihara,Ohno2,Oiwa,Dietl}. Spin injection from ferromagnetic (Ga,Mn)As into nonmagnetic semiconductors \cite{YOhno} and the large tunneling magnetoresistance (TMR) of (Ga,Mn)As-based magnetic tunnel junctions \cite{Tanaka,Chiba} are typical examples associated with such interesting properties. To further develop micro- or nano-sized spin electronic devices using ferromagnetic semiconductors, novel techniques for modifying the ferromagnetic properties of (Ga,Mn)As on the micro- or nano-meter scale are required.

Since the ferromagnetism of (Ga,Mn)As is sensitive to hole carrier concentration, a modification in the ferromagnetism of (Ga,Mn)As is expected to be achieved, given a systematic variation in the carrier concentration in (Ga,Mn)As. Ga$^{+}$ ion irradiation is one of the most suitable tools for modifying the carrier concentration in GaAs on the micro- or nano-meter scale: this technique has been used to control the conductivity of GaAs and to fabricate mesoscopic semiconductor structures.\cite{Hirayama}  Being motivated by the prospect of controlling the carrier concentration in (Ga,Mn)As with this approach, we have recently reported a preliminary experiment on the influence of Ga$^{+}$ ion irradiation on the ferromagnetic properties of (Ga,Mn)As using a focused ion beam (FIB) technique.\cite{Kato} However, a systematic variation in the ferromagnetic features of (Ga,Mn)As was not obtained, although a sure sign that Ga$^{+}$ ion irradiation altered the hole carrier concentration and the corresponding magnetic properties was detected.

In this study, we demonstrate for the first time the systematic control of ferromagnetism such as magnetic anisotropy and Curie temperature in $p$-type (Ga,Mn)As by adjusting Ga$^{+}$ ion irradiation dosage. This technique can be applied to fabricating geometrically modified magnetic structures on the nanometer scale in a single (Ga,Mn)As epilayer.
\begin{figure}[b]
\begin{center}
\includegraphics[width=8cm]{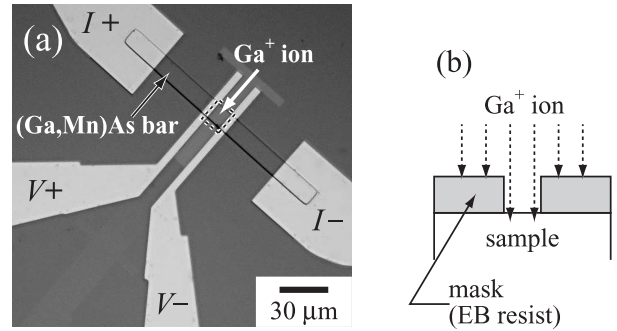}
\caption{(a) Optical micrograph of (Ga,Mn)As bar structure irradiated with Ga$^{+}$ ion beam. (b) Schematic diagram of ion irradiation process for sample covered with EB resist as mask.}
\end{center}
\end{figure}

A Ga$_{0.949}$Mn$_{0.051}$As epilayer with a thickness of 75 nm was grown on a GaAs/GaAs (001) substrate using molecular beam epitaxy (MBE), followed by low-temperature annealing at 245$^{\circ}$C in N$_2$ gas atmosphere for 240 min.\cite{Kato2} The epilayer was patterned into 10-$\mu$m-wide bar structures (10$\times$150 $\mu$m$^{2}$) with Ti/Au ohmic contacts using electron beam (EB) lithography and wet etching for transport measurements. The long axis of the samples was oriented along GaAs [100] [Fig. 1(a)]. The samples were subjected to ion irradiation using a 30 keV Ga$^{+}$ FIB (FB-2000A, Hitachi). The ion irradiation conditions were controlled by adjusting the beam current ranging from 4 to 40 pA and the irradiation area region from 10$^{2}$ to 10$^{4}$ $\mu$m$^{2}$ under a constant irradiation time of 12 s ; the irradiation dosage was varied from 4.0 $\times$ 10$^{12}$ to 4.0 $\times$ 10$^{15}$ ion/cm$^{2}$ in the 10 $\times$10 $\mu$m$^{2}$ area located in the region between two voltage electrodes [see Fig. 1(a)]. We have confirmed using atomic force microscopy that the thickness of an etched layer subjected to the ion irradiation is less than 2 nm. To protect the voltage electrodes from being damaged by ion irradiation, the samples except for the area to be irradiated were masked with EB resist  [Fig. 1(b)]. Our calculation suggests that Ga$^{+}$ ions penetrate down to several ten nm underneath the EB resist so no damages occur in (Ga,Mn)As covered with EB resist. Transport measurements were performed by means of a standard four-probe method using  a physical property measurement system (PPMS, Quantum Design), and magnetic fields were applied parallel to the long axis of the bar structure ([100]). 

\begin{figure}[t]
\begin{center}
\includegraphics[width=8cm]{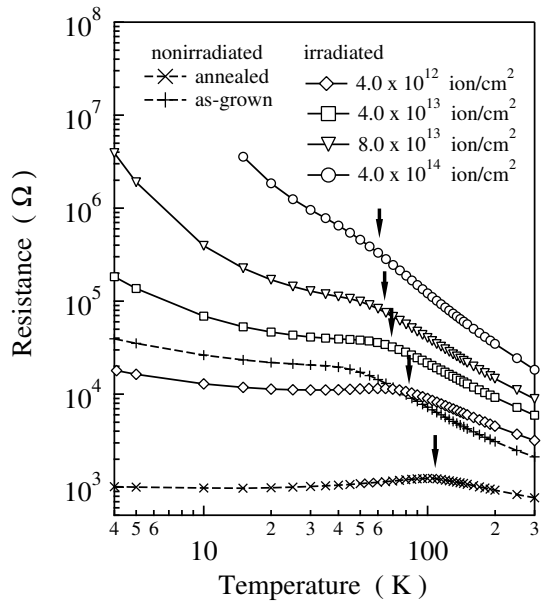}
\caption{Temperature-dependent resistances of (Ga,Mn)As bars subjected to Ga$^{+}$ ions at various irradiation dosages.}
\end{center}
\end{figure}

The temperature-dependent resistances of (Ga,Mn)As samples are shown in Fig. 2 as a function of Ga$^{+}$ ion irradiation dosage. The data of nonirradiated as-grown and annealed samples are also shown as references. Both nonirradiated samples show clear metal-insulator transitions near the Curie temperature ($T_{\textnormal{c}}$),\cite{Matsukura} and the annealing decreases their resistance by a factor of 1/30 at 4 K.\cite{Kato2,Edmonds} We define as $T_{\textnormal{p}}$ the metal-insulator transition temperature. $T_{\textnormal{p}}$ is also increased by a factor of 2 by annealing, indicating that annealing leads to an increase in the hole carrier concentration.\cite{Kato2,Edmonds} The annealed samples were then irradiated with the Ga$^+$ ion beam at various irradiation dosages using an FIB technique. With increasing irradiation dosage, $T_{\textnormal{p}}$ shifts toward a lower temperature and resistance increases up to $\sim$M$\Omega$. In particular, the insulating behavior at a low temperature becomes significant with increasing irradiation dosage. 
\begin{figure}[t]
\begin{center}
\includegraphics[width=8cm]{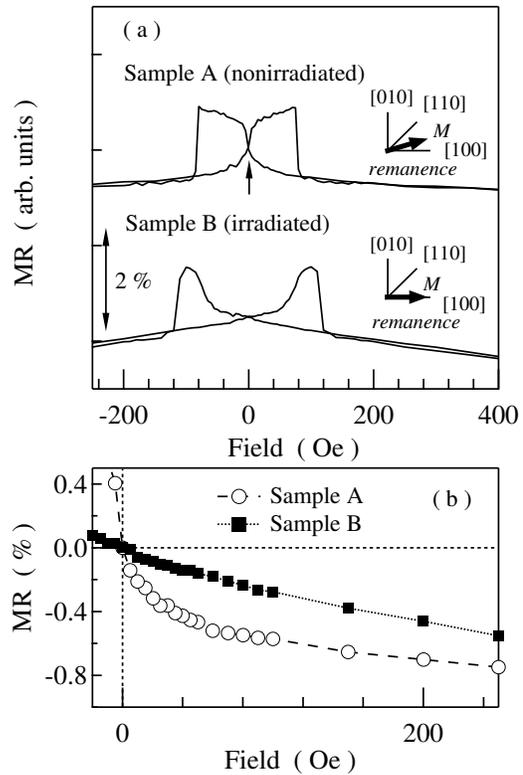}
\caption{(a) MR hysteresis curves of samples A and B at 4 K and (b) enlargement of MR curves near zero field.}
\end{center}
\end{figure}

Figures 3 show the field-dependent resistances (MR curves) of a nonirradiated annealed sample (sample A) and the corresponding irradiated sample (sample B) with an ion irradiation dosage of  $4.0 \times 10^{12}$ ion/cm$^2$  at 4 K. As seen in Fig. 3(a), an increase in MR is observed at remanence for sample A as indicated by an arrow; on the other hand, no visible changes in MR are seen at remanence for sample B. This feature is more clearly seen in an enlargement of the low-field regime from +250 Oe to $-$20 Oe [Fig. 3(b)]: the MR change for sample A is larger than that for sample B. Because the magnetoresistance of (Ga,Mn)As  is generally explained in terms of anisotropic magnetoresistance (AMR)\cite{Hamaya2,Kato2}, we can deduce the orientation of magnetization from the MR feature: the magnetization of sample A tilts from [100] toward [110] while that of sample B remains along [100] at remanence. Schematic illustrations of the magnetization direction at remanence are depicted in the inset of Fig. 3(a) for both samples. The results provide clear evidence that the contribution of [110] uniaxial anisotropy for the irradiated sample is smaller than that for the nonirradiated sample. 

In our recent studies,\cite{Kato2,Hamaya1} it has been found that the in-plane magnetic anisotropy of (Ga,Mn)As/GaAs(001), which is composed of both $\langle100\rangle$ cubic magnetocrystalline anisotropy and [110] uniaxial anisotropy,\cite{Hamaya1,Hamaya2,Welp} strongly depends on hole concentration, and that the contribution of [110] uniaxial anisotropy is enhanced with increasing hole concentration.\cite{Hamaya1,Kato} 
Thus, it is likely that irradiated samples, which show a small contribution of [110] uniaxial anisotropy, have a lower hole concentration than noniraddiated samples.\cite{Hamaya1,Kato} To obtain clearer information about the mechanism of the change in the transport properties and the magnetic anisotropy shown in Figs. 2 and 3, we estimate the hole concentration of irradiated samples from the $T_{\textnormal{c}}$ vs hole concentration data shown in the inset of Fig. 4, assuming $T_{\textnormal{c}}$ $\approx$ $T_{\textnormal{p}}$. Because the Curie temperature and magnetic anisotropy of (Ga,Mn)As correlate with hole concentration\cite{Dietl,abolfath}, we expect a close relationship between hole concentration and ion irradiation dosage. Figure 4 shows the hole concentration $p$ (left axis) of irradiated samples as a function of Ga$^+$ ion irradiation dosage. A systematic decrease in hole concentration is seen with increasing ion irradiation dosage, indicating that Curie temperature and magnetic anisotropy can be contolled by adjusting the dosage.
\begin{figure}
\begin{center}
\includegraphics[width=8cm]{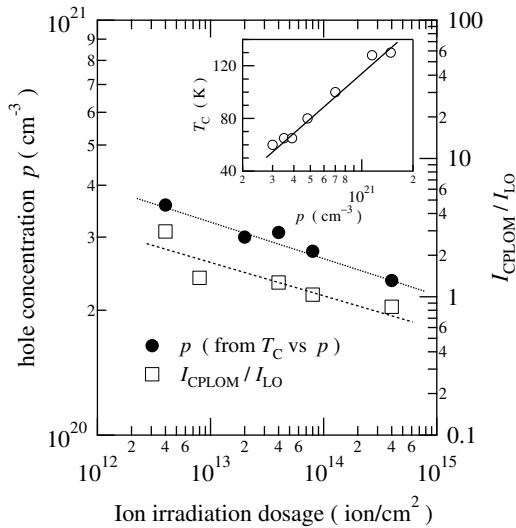}
\caption{Hole concentration $p$ (left axis) and ratio of peak intensities $I_{\textnormal{CPLOM}} / I_{\textnormal{LO}}$ (right axis) as a function of Ga$^{+}$ ion irradiation dosage. The inset shows the relation between $T_{\textnormal{c}}$ and $p$ measured by the electrochemical capacitance-voltage (ECV) method \cite{moriya} for various (Ga,Mn)As samples.}
\end{center}
\end{figure}

Micro-Raman data for various irradiated samples with different dosages also corroborate our proposal of the effect of ion irradiation.\cite{Kato,seong} Detailed procedures of the Raman measurements and analyses are described elsewhere.\cite{Kato} 
We note that the Raman data exhibit no influence of structural disorder in irradiated samples because the data were collected after removing 25 nm of (Ga,Mn)As from the surface using a wet etching technique and the crystal structure of the remaining (Ga,Mn)As was confirmed to be the same as that of the nonirradiated samples by high-resolution X-ray diffraction analysis.\cite{Kato} An essential feature of the Raman spectra is the following: the spectra consist of two peaks which originate from a longitudinal optical (LO) mode and a coupled plasmon longitudinal optical mode (CPLOM) overlapped with a transverse optical (TO) mode.\cite{Kato,seong,irmer} We plot $I_{\textnormal{CPLOM}}$$/$$I_{\textnormal{LO}}$ (right axis) as a function of Ga$^+$ ion irradiation dosage in Fig. 4. A clear irradiation dosage dependence of  $I_{\textnormal{CPLOM}}$$/$$I_{\textnormal{LO}}$ is observed similar to that of hole concentration, which is consistent with the fact that CPLOM is associated with carrier concentration. Therefore, we conclude that the changes in $T_{\textnormal{p}}$ (Fig. 2) and magnetic anisotropy (Fig. 3) with Ga$^+$ ion irradiation are due to a reduction in hole concentration.\cite{Dietl,Hamaya3,Kato2} Also, the reduction in hole concentration may be attributed to the formation of deep trap levels, as we proposed in our previous report.\cite{Kato} 

In summary, we have demonstrated the control of the ferromagnetism of (Ga,Mn)As using Ga$^+$ FIB. Curie temperature and magnetic anisotropy can be controlled by adjusting Ga$^+$ ion dosage as seen from magnetoresistance and micro-Raman data, which we attribute to a change in hole concentration due to the formation of deep trap levels. Because the FIB technique has an advantage of focusing a beam on the nanometer scale, we expect it to be utilized as a nanoscale modification tool for determining the magnetic properties of (Ga,Mn)As with the prospect of developing novel spin electronic devices.

\section*{Acknowledgment}
The authors gratefully thank Professor Y. Yamazaki of the Tokyo Institute of Technology (T. I. Tech.) for kindly offering the opportunity to use their FIB. Stimulating comments on Raman spectra by Professor K. G. Nakamura of T. I. Tech. are greatly appreciated. The authors also thank Dr. T. Kondo of T. I. Tech. for his experimental support and useful discussion.

\end{document}